# Status of MUSIC, the MUltiwavelength Sub/millimeter Inductance Camera

Sunil R. Golwala[*a], Clint Bockstiegel[b], Spencer Brugger[b], Nicole G. Czakon[a], Peter K. Day[c], Thomas P. Downes[a], Ran Duan[a], Jiansong Gao[d], Amandeep K. Gill[b], Jason Glenn[b], Matthew I. Hollister[a], Henry G. LeDuc[c], Philip R. Maloney[b], Benjamin A. Mazin[e], Sean G. McHugh[e], David Miller[a], Omid Noroozian[d], Hien T. Nguyen[c], Jack Sayers[a], James A. Schlaerth[a], Seth Siegel[a], Anastasios K. Vayonakis[a], Philip R. Wilson[c], Jonas Zmuidzinas[a,c]

[a]Division of Physics, Mathematics, and Astronomy, California Institute of Technology, Pasadena, CA, USA, 91125;
[b]CASA, University of Colorado, UCB 593, Boulder, CO, USA, 80309;
[c]Jet Propulsion Laboratory, 4800 Oak Grove, Pasadena, CA, USA, 91109;
[d]Quantum Sensors Group, National Institute of Standards and Technology, Boulder, CO, USA, 80305;
[e]Department of Physics, University of California, Santa Barbara, CA, USA, 93106

## ABSTRACT

We present the status of MUSIC, the MUltiwavelength Sub/millimeter Inductance Camera, a new instrument for the Caltech Submillimeter Observatory. MUSIC is designed to have a 14', diffraction-limited field-of-view instrumented with 2304 detectors in 576 spatial pixels and four spectral bands at 0.87, 1.04, 1.33, and 1.98 mm. MUSIC will be used to study dusty star-forming galaxies, galaxy clusters via the Sunyaev-Zeldovich effect, and star formation in our own and nearby galaxies. MUSIC uses broadband superconducting phased-array slot-dipole antennas to form beams, lumped-element on-chip bandpass filters to define spectral bands, and microwave kinetic inductance detectors to sense incoming light. The focal plane is fabricated in 8 tiles consisting of 72 spatial pixels each. It is coupled to the telescope via an ambient-temperature ellipsoidal mirror and a cold reimaging lens. A cold Lyot stop sits at the image of the primary mirror formed by the ellipsoidal mirror. Dielectric and metal-mesh filters are used to block thermal infrared and out-of-band radiation. The instrument uses a pulse tube cooler and $^3$He/$^3$He/$^4$He closed-cycle cooler to cool the focal plane to below 250 mK. A multilayer shield attenuates Earth's magnetic field. Each focal plane tile is read out by a single pair of coaxes and a HEMT amplifier. The readout system consists of 16 copies of custom-designed ADC/DAC and IF boards coupled to the CASPER ROACH platform. We focus on recent updates on the instrument design and results from the commissioning of the full camera in 2012.

## 1. INTRODUCTION

The past decade has seen the advent of kilopixel arrays in submillimeter- and millimeter-wave astronomy, enabling the mapping of large areas of sky at these wavelengths for the first time. Such arrays have detected galaxy clusters out to their redshift of formation, resolved a significant portion of the cosmic far-infrared background into dusty, star-forming galaxies (DSFGs), surveyed for protostellar cores and young stellar objects in our own galaxy, and mapped the cosmic microwave background's temperature and polarization anisotropy. They have opened up the submillimeter/millimeter-wave sky and revolutionized the way we understand the formation of stars, galaxies, and clusters of galaxies as well as measure the fundamental parameters of our universe.

A key enabling technology for these developments has been multiplexable superconducting detectors in the form of time-domain- and frequency-domain-multiplexed transition-edge sensors (TESs). Future telescopes, in particular

---

[*]golwala@caltech.edu

CCAT[1,f], will provide fields-of-view of order a square degree combined with few-arcsecond resolution, requiring hundreds of thousands, millions, and even tens of millions of detectors for their full exploitation. To reach these enormous detector counts requires another revolution in multiplexable detectors. Microwave kinetic inductance detectors (MKIDs) are one attractive option for such focal plane arrays because of their high multiplex density relative to the above TES architectures.

Another enabling technology has been the development of mass-producible detector architectures. Kilopixel arrays have required wafer-scale detector production, mass production of feed arrays, and large-aperture quasioptical filters. An important next step in this direction is fully photolithographic feed arrays and bandpass filters.

The development of the MUltiwavelength Sub/millimeter Inductance Camera (MUSIC) has been motivated by the aforementioned compelling science and the availability of these new technologies. MUSIC will demonstrate a fully photolithographic focal plane using broadband slot-dipole phased-array antennas for beam definition, bandpass filter banks for mm-wave spectral band selection, and MKIDs for power detection. It will have one of the largest detector counts of any submillimeter/millimeter-wave imaging array (2304) and the broadest simultaneous spectral coverage at this array size (four bands). MUSIC is a natural pathfinder for more ambitious future instruments such as CCAT LWCam.[1]

## 2. SCIENCE GOALS

MUSIC has been designed with the goal of high mapping speed and high angular resolution in multiple submm/mm spectral bands relevant for the study of the Sunyaev-Zeldovich (SZ) effect in galaxy clusters, dusty star-forming galaxies, and protostellar and young stellar cores in our own galaxy.

The SZ effect enables the study of the pressure and total thermal energy in the intracluster medium (ICM) in galaxy clusters. Questions of particular interest include: Is there a universal pressure profile for galaxy clusters, and over what radial range relative to the cluster diameter? How does this pressure profile depend on the accretion, merger, and galaxy feedback history of the cluster? To what extent are clusters truly virialized? What is the detailed mechanism of the shocks by which kinetic energy of accreting material is converted to random thermal energy? Are there residual bulk motions in galaxy clusters and what do these tell us about their formation history? How important are the contributions of turbulence, magnetic fields, and cosmic rays to the overall pressure equilibrium of galaxy clusters? Cosmological studies are also of interest, including: What is the mass function of galaxy clusters as a function of redshift and what does this tell us about cosmological parameters? What is the secondary anisotropy spectrum due to galaxy clusters and what does this tell us about the nature of pressure profiles of galaxy clusters? What is the time duration and detailed history of reionization as reflected in the amplitude and shape of the power spectrum of kinetic SZ anisotropy? To answer the first set of questions requires an instrument capable of mapping massive galaxy clusters out to the virial radius and in multiple spectral bands to enable the separation of cluster thermal and kinetic SZ effects from primordial CMB anisotropy, DSFGs, and radio point sources. A field of view of 10' to 20' is required combined with sub-arcminute angular resolution and at least three and preferably more spectral bands. The latter set of questions requires, in addition, high mapping speed and large field-of-view.

High-redshift, dusty, star-forming galaxies are a class of objects largely undiscovered prior to the mid-1990s yet which emit roughly half of the energy density liberated by stellar nucleosynthesis over cosmic time. Their place in the paradigm of galaxy formation and evolution is understood in broad strokes—they reflect an epoch of extremely intense star-formation in the $z > 1$ universe when large gas reservoirs and merger activity drove star-formation rates enormously larger than in the present-day universe, creating very dusty objects whose stellar photon emission was largely reprocessed into the far-infrared and submillimeter—but many details remain to be understood. At roughly 1 mm wavelength, the negative K-correction for dusty star-forming galaxies[2] renders the observed flux density an essentiallly redshift-independent measure of the DSFG luminosity. Thus, mm-wave surveys for DSFGs can find such objects out to any redshift at which they exist. Combining information from mm-wave bands with submm bands (e.g., from *Herschel* SPIRE) enables the selection of the highest-redshift objects at $z > 4$. The study of the number counts and clustering properties of these galaxies can reveal their connection to the underlying dark matter halo distribution and can tie them into models of galaxy formation. Discovering these objects requires a millimeter-wavelength imager with fast mapping speed. Beginning to understand the connection to dark matter halos requires multiple wavelengths for the study of clustering functions at individual wavelengths and across wavelengths. Multiwavelength data can also constrain the dust emissivity spectral index and reveal the nature of the dust grains in these very dusty objects.

---

[f]http://www.ccatobservatory.org

A complete census of protostellar cores and young stellar objects in our own galaxy would provide critical input to models of star formation. A detailed understanding of star formation would predict the mass function and clustering properties of these stellar nurseries. Thus, an instrument capable of finding even the coldest cores over many hundreds of square degrees of sky would have substantial impact. Multiwavelength data could be used to refine models of the dust grains providing the emission and thus better relate observed fluxes to underlying dust and gas masses.

MUSIC has been designed with these goals in mind. To enable fast mapping, it uses the entire physically available field-of-view of the Caltech Submillimeter Observatory, which provides 23" to 45" angular resolution in the MUSIC bands. It deploys a large detector count in a mass-producible focal plane using the aforementioned technologies. Its design has the goal of approaching background-limited performance to maximize mapping speed. And it provides uniquely broad wavelength coverage to maximize the astrophysical information available.

## 3. INSTRUMENT DESIGN AND MEASURED PERFORMANCE

### 3.1 Optical Train

The optical design of MUSIC was described in detail elsewhere.[3] Here, the design criteria and design are reviewed and the on-sky performance is presented.

#### 3.1.1 Design Criteria

MUSIC was designed to make use of the maximum field-of-view available at the CSO, limited by the hole in the primary mirror and support structure through which the optical path passes. A requirement of < 1% vignetting by these obstructions leads to a 15' diameter field-of-view, motivating a square focal plane 14' on a side.

The choice of mm-wave bands—0.87, 1.04, 1.33, and 1.98 mm—was driven by the science goals, as described in Section 2. Given these bands and the field-of-view, the next design variable was the pixel size. For the antenna-coupled technology that provides four simultaneous mm-wave bands, the pixel size is the same at all wavelengths, and so the choice of pixel size must be a compromise between the different science goals and the individual bands' optimizations. Individual pixel sensitivity drives one toward pixels approaching $2(f/\#)\lambda$ in size because of the high optical efficiency thus obtained. Angular resolution, which is critical for penetrating deeper into the DSFG population by obtaining as low a confusion limit as possible, drives toward smaller pixels that illuminate the primary mirror more uniformly. Also, a smaller pixel size yields a larger number of pixels, and, if the individual pixels are background-limited, then the mapping speed is either independent of (shot noise limit) or increases linearly (Bose noise limit) with the number of pixels, while it decreases linearly with the number of pixels if detector- or readout-noise limited. The pixels were therefore made as large as possible without substantially degrading the resolution in the 0.87-mm band relative to the geometrical limit of the telescope for a 9.0-m diameter illumination, 20". (This illumination of the 10.4-meter primary mirror was justified elsewhere.[3]) The final pixel parameters are given in Table 1.

#### 3.1.2 Optical Design, Implementation, and Performance

The detailed design of the optical train, in particular the physical constraints on it, has been described elsewhere.[3] The CSO's $f/12.6$ Cassegrain focus feeds into relay optics consisting of two flat mirrors (to fold the beam), an ellipsoidal tertiary mirror, and a cold lens, as shown in Figure 1. The ellipsoidal mirror and lens parameters were chosen to maximize the image quality and telecentricity over the field-of-view desired. The parameters of the mirror and the lens are also listed in Figure 1. The lens and focal plane are displaced by 5 mm in elevation to obtain improved telecentricity at an acceptable cost in Strehl ratio. A calculation of the resulting Strehl ratio as a function of position off-axis is shown in Figure 1. There is a clear asymmetry in the Strehl ratio with elevation because the ellipsoidal mirror breaks the symmetry. A physical optics calculation in Zemax was used to ensure spillover past any mirrors to warm surfaces would be < 1%, while spillover past the secondary was allowed to be about twice as large because of the background sky's low emissivity.

Table 1: Parameters of MUSIC optical design. The pixel size is relative to the focal ratio at the focal plane, which has been increased by the Lyot stop. The individual efficiency contributions are calculated, not measured, values. "Phonon" is the efficiency for conversion of optical photon power to quasiparticles; for $h\nu \gg 2\Delta$, it asymptotes to 0.58,[17] but it must allow creation of two quasiparticles per phonon for $h\nu \geq 2\Delta$, requiring it to increase to unity as $h\nu \to 2\Delta$. "Antenna" is the radiative efficiency of the antenna (see Figure 4). The "Filters" efficiency incorporates loss and reflections at the dielectric, metal-mesh, and infrared shader filters. The "Mirrors" efficiency consists of approximately 1% at each surface due to imperfect reflectivity (five surfaces) along with a contribution from Ruze scattering at the primary mirror (which degrades the efficiency more at shorter wavelengths). The "Measured" efficiency is the range over the four subarrays (two tiles) that have been characterized in detail.

| Band | λ [mm] | Δν [GHz] | Pixel size (f/#)λ | FWHM ["] | Optical efficiency | | | | | | | |
|------|--------|----------|-------------------|----------|---------|-------------|---------|-----------|---------|---------|-------|----------|
|      |        |          |                   |          | Phonon  | Micro-strip | Antenna | Lyot stop | Filters | Mirrors | Total | Measured |
| B0   | 1.98   | 34       | 0.73              | 44       | 0.66    | 0.88        | 0.74    | 0.28      | 0.74    | 0.94    | 0.084 | 0.067–0.074 |
| B1   | 1.33   | 45       | 1.08              | 31       | 0.58    | 0.83        | 0.70    | 0.51      | 0.76    | 0.94    | 0.12  | 0.10–0.12 |
| B2   | 1.04   | 34       | 1.39              | 25       | 0.58    | 0.79        | 0.78    | 0.68      | 0.69    | 0.94    | 0.16  | 0.075–0.083 |
| B3   | 0.87   | 21       | 1.66              | 23       | 0.58    | 0.75        | 0.71    | 0.76      | 0.61    | 0.91    | 0.13  | 0.032–0.091 |

The relay optics and internal structure of the cryostat were designed to ensure deflections due to gravity and accelerations were kept subdominant to the telescope's intrinsic pointing scatter (as opposed to deflection, which occurs with zenith angle and is reproducible and corrected for automatically.) The CSO itself is capable of pointing accuracy at the 2" to 3" level, as demonstrated by observations with the SHARC-II and ZEUS 350-μm instruments, which reside at the Nasmyth focus and thus themselves do not contribute to pointing scatter. This goal has been achieved, as shown in Figure 2, where a rms pointing scatter of approximately 3" is demonstrated relative to the intrinsic telescope pointing model.

The dielectric elements in the optical design, consisting of filters and the cold lens, are shown in the detailed layout of the dewar optics in Figure 1. The vacuum window is HDPE, with a diameter of 375 mm, a clear diameter of 300 mm, and a thickness of 12.6 mm. Because of HDPE's good thermal conductivity and the large thickness of the window, even the central portion of the window facing vacuum remains quite warm and thus presents a large blackbody load at frequencies at which it becomes emissive, above ~ 1 THz. To reduce the thermal infrared radiation load from this window, it is followed by two metal-mesh IR shaders (spares from the SPIDER CMB polarimeter) produced by the Cardiff University group;[4] one with 20-μm (15 THz) cutoff wavelength and one with 80-μm (3.75 THz) cutoff wavelength. These sit at the PT1 (roughly 45 to 60 K) stage of the dewar. While they filter a good deal of the 300 K blackbody radiation, PTFE filters at the PT1 and PT2 (roughly 4 K) stages of the dewar are nevertheless required. The PT1 filter consists of two PTFE slabs of thickness 15.6 mm. The first slab heats up appreciably in the center, to 90 K, which is hot enough that the PTFE begins to radiate substantially in its emissive band above 1 THz. The second PTFE filter at the PT1 stage absorbs much of that radiation yet remains isothermal at 60-K temperature of the PT1 radiation shield lid because the load on it is much less than the load on the first filter. A single 15.9-mm thick, 250-mm clear-aperture piece of PTFE is used at PT2; because the load it must absorb is much smaller, it never heats enough to radiate above 1 THz. Inside the PT2 shield sits the optics tube, which has the Lyot stop at one end, the focal plane at the other, and the cold lens in the center. A 420 GHz low-pass edge metal-mesh filter (QMC Instruments, not antireflection-coated) sits at the Lyot stop. The 420 GHz filter is placed here, rather than at the PT2 shield, because its primary purpose is to limit the bandwidth and the optical load on the sub-Kelvin cooler, the bulk of whose surface area (the focal plane array, roughly a square 135 mm on a side) is inside the optics tube. With the filter, the load on the focal plane is 1 μW; without it, the load would be 10 μW. The low-pass filter also serves to limit the mm-wave bandwidth the detectors see, critical for minimizing direct absorption in the MKIDs. The lens is made of UHMWPE and resides midway down the optics tube. HDPE and UHMWPE were chosen for the window and lens because they offered a reasonable tradeoff between loss, cost, and the challenge of antireflection coating given the large diameters, with HDPE chosen for the vacuum window because of its greater strength. All the dielectric elements have been anti-reflection

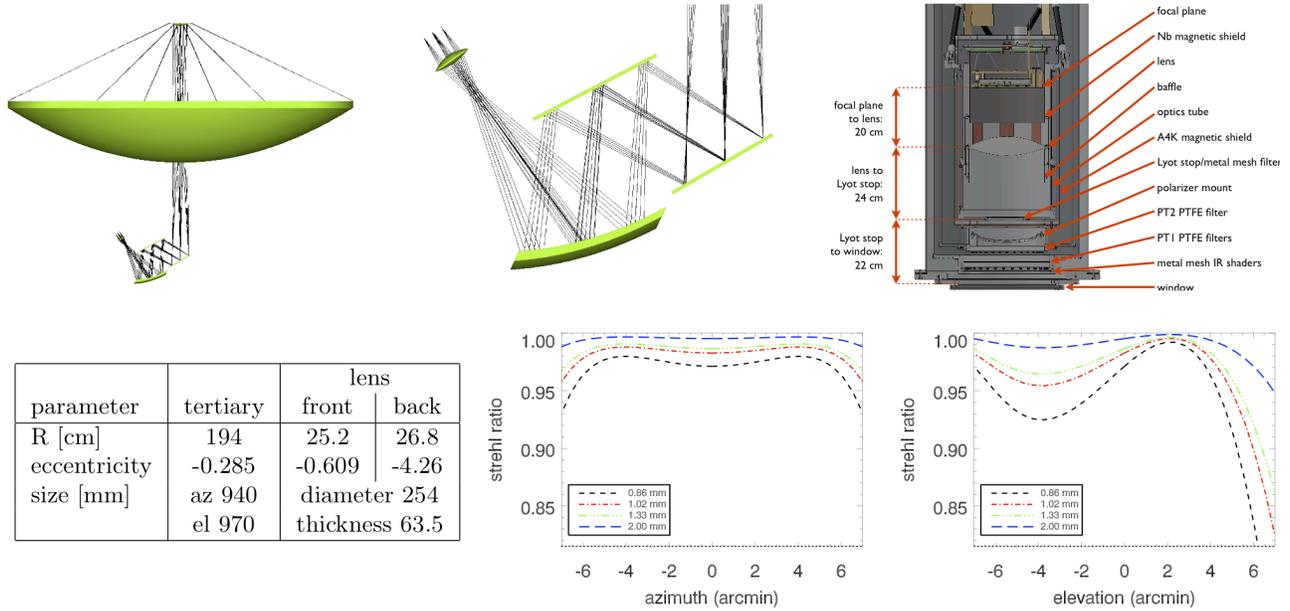

Figure 1: MUSIC optical design. Top left and center: geometrical optics ray trace. Top right: cross section of the portion of the dewar between the FPU and the vacuum window, illustrating the cold optics and dielectric filters. Bottom left: Parameters of the ellipsoidal tertiary mirror and the cold lens. Bottom right: Calculated Strehl ratio as a function of coated on all dielectric/vacuum interfaces with single layers of 250-μm Porex PM23J ($n$ = 1.23) with LDPE glue layers. Its refractive index is well-matched to HDPE, UHMWPE, and PTFE. The overall transmission of the dielectric elements due to loss and reflections is given in Table 1.

The geometrical optics performance has been tested during the MUSIC commissioning run in May, 2012. Figure 2 shows a histogram of beam FWHMs by band and Figure 3 shows a map of the focal plane projected onto the sky. The map consists of the full range of pixels in azimuth and 25% of the range in elevation (the portion of the array that was populated for this run). The asymmetry in the azimuthal direction is due to a rotation of the dewar by 4° relative to alignment of the rows with the azimuth direction, which yields more uniform sky coverage in a single scan. As the histogram shows, the beam FWHMs have been verified, but the map indicates the distortion pattern deviates appreciably from expectations. While image quality in the form of FWHM appears unaffected and thus this distortion does not limit the instrument performance, the source of the mismatch will nevertheless be pursued.

Stray light is a critical challenge in instruments such as MUSIC, and measures have been taken to minimize it. While dielectric filters such as PTFE have nonnegligible loss, they do not suffer from high-angle scattering of high-frequency radiation as may be the case with metal-mesh filters.[5] The only critical mesh filter is deep inside the system, at the Lyot stop, and thus, even at high angles, it sees relatively cold surfaces—the inside of the PT2 radiation shield or, worst case, the inside of the PT1 radiation shield. However, in the time-reversed sense, there is substantial optical throughput terminating on the Lyot stop, as indicated in Table 1. If this throughput is not fully absorbed, multiple reflections can lead to its escape at high angles through the optical path. Rays terminating just outside the edge of the Lyot stop can escape with two bounces, one off the Lyot stop and one off the focal plane. For B0, whose Lyot stop throughput is 28%, only 3% of the entire beam (4% of the nominally terminated beam) would have to be reflected to result in a 10% contribution to the B0 beam passing through the Lyot stop from this reflected component. The inside of the optics tube, the baffle extending above the lens, and the Nb magnetic shield (Section 3.3) have been blackened with Stycast loaded with carbon[6] and stainless steel[7,†] powders troweled to have V-shaped grooves to absorb this throughput.

There is, however, evidence that these surfaces are not sufficiently absorptive: 20%, 13%, 10%, and 9% (B0, B1, B2, B3) of the beam exiting the dewar spills past the ellipsoidal mirror onto warm surfaces in the optics box; see Figure 2.[‡]

---

†Carpenter 316L Micro-Melt powder, mixed to 7% by weight, donated by Carpenter Technology Corporation.
‡The numbers are obtained by comparing the difference in power received between 300K and 77K loads presented at the dewar window to 300K and 77K loads presented in place of the first folding flat mirror. This does not test for spillover past that flat and at

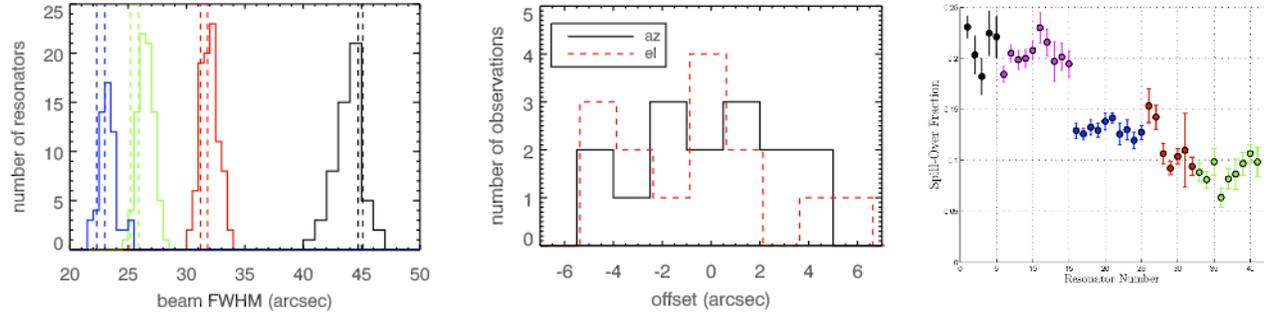

Figure 2: MUSIC optical performance. Left: Measured beam FHWMs, by mm-wave band. The color key to the mm-wave bands is: B0: black; B1: red; B2: green; B3: blue. The expected range of values (varying with position on the focal plane) are indicated with dashed vertical lines. There is reasonable agreement, though some systematic deviations may be apparent. Center: MUSIC pointing scatter relative to CSO pointing model based on observations of Mars above elevation of 25°, yielding rms scatters of approximately 2.5" in azimuth and 3.5" in elevation. The only pointing correction made is the CSO's own pointing model. The elevation histogram is slightly displaced to the right for visibility. The achieved rms are already better than was achieved with Bolocam *after* correcting for observed variations of pointing with azimuth and elevation, indicating the much improved stiffness of the MUSIC optics box. Right: Spillover fraction in the optics box for a sample of detectors. The first five, black data points are for nominally "dark" resonators, with a mean of 22% spillover. The resonators then proceed in the order B0, B1, B2, B3 with mean spillover fractions of 20%, 13%, 10%, 9%.

In the future, an internal baffle, consisting of a set of concentric blackened tubes extending from the Lyot stop toward the lens, will be installed to prevent such reflections.

### 3.2 Detectors

#### 3.2.1 Optical Coupling and Millimeter-Wave Band Definition

Optical power is received at the focal plane by superconducting slot-dipole phased-array antennas, the design of which is described elsewhere.[8,9] Briefly, the antenna consists of 32 slot dipoles in a 150-nm thick ground plane, each tapped in 32 places. The slots are 18 μm wide and 4.2 mm long. The taps form the beginning of a binary summing tree in microstrip, combining first all the power from a single slot and then combining the power from the various slots, all coherently. The microstrip layers are the ground plane, 400 nm of silicon nitride, and 450 nm of niobium. The dielectric and top layers of the microstrip cross the slot and terminate in a capacitor to provide a short to the ground plane. An electric field excited across the slot (polarization perpendicular to the slot) creates a voltage on the microstrip tap. The microstrip begins with width 1 μm at the slots, exits the first stage of the combiner tree with width 2 μm, tapers to 1.7 μm at the entrance to the next stage of the combiner, exits that stage with width 4 μm, and then repeats the taper/expansion pattern through the tree to its exit. The effective electric field excitation pattern is well approximated by uniform illumination of the 4.2 mm × 4.2 mm pixel. The antenna resides on a silicon substrate, through which it is illuminated.

The bandwidth of the antenna is set by the tap spacing and the slot length. The grating lobe defined by the tap spacing must be beyond the horizon for the antenna to efficiently radiate into its main beam, which is defined by the condition $\lambda/n > \delta$ where $\delta$ is the tap spacing, $\lambda$ is the free-space wavelength, and $n$ is the index of refraction of the dielectric substrate (approximately 3.5 for silicon). The low-frequency edge of the band is the frequency for which the guide wavelength in the slot approaches the slot length, at which point slot-end effects dominate and the impedance deviates from a good match to the microstrip taps.

Figure 4 shows the calculated radiative efficiency of the antenna as a function of frequency when fabricated on a silicon wafer of thickness 369 μm and mounted with a 132 μm z-cut crystal quartz anti-reflection layer (on the opposite side of the silicon as the antenna) and a backshort placed on the vacuum side of the antenna at a spacing of 250 μm. These

---

the entrance aperture of the optics box, but the bulk is expected to be spillover outside the ellipsoidal mirror because the ensuing flat mirrors are oversized relative to it.

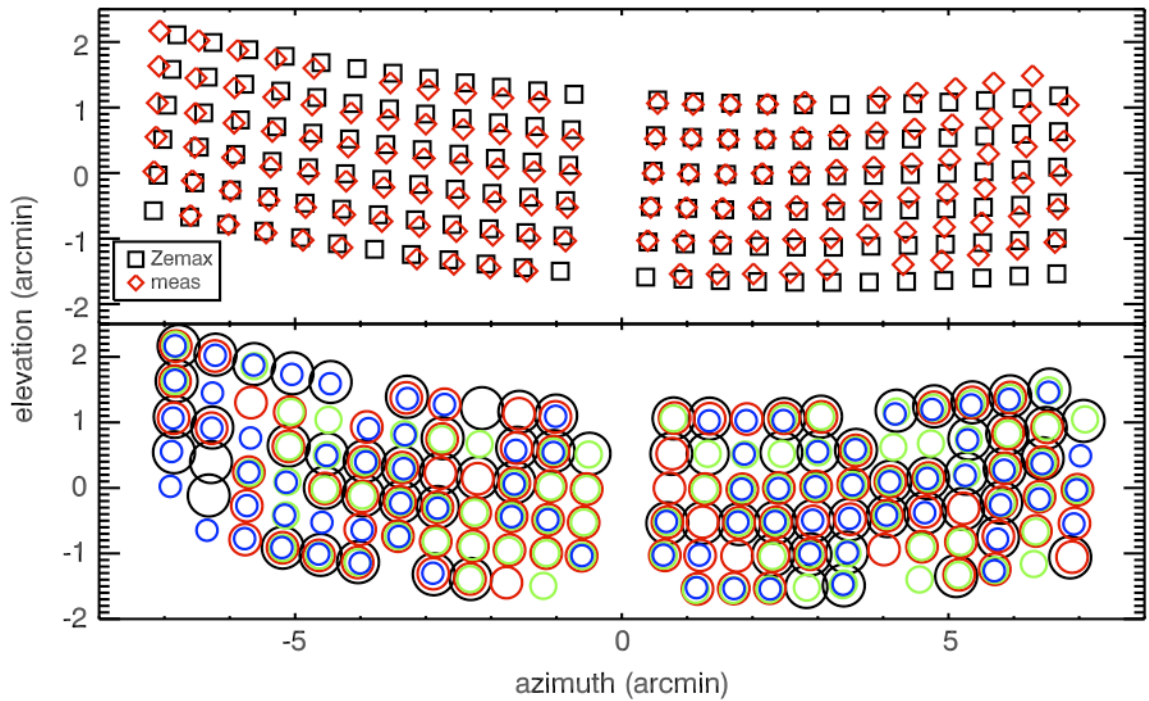

Figure 3: MUSIC optical performance. A map projected onto the sky of the two subarrays (25% of the full focal plane) employed during the May, 2012, commissioning run based on maps of bright point sources such as Uranus. The top plot compares the observed and expected positions of the beam centroids. There is an as-yet unexplained deviation from the expectation, larger in the positive azimuth direction. The bottom plot shows circles centered on the reconstructed positions with diameters indicating the reconstructed beam FWHM and colors identifying the mm-wave bands (see Figure 2). In the first analysis of these data and of the 144 pixels possible, 26 pixels have all four optical bands functional, and 44 have three out of four bands functional.

dimensions were chosen to provide roughly equal radiative efficiencies in the four mm-wave bands, indicated approximately in Figure 4.

The antenna feed connects to a bank of bandpass filters, shown in Figure 5. The microstrip tapers to 2 μm width as it enters the filter bank. The filters use a 3-pole Butterworth design, transformed to use no shunt inductances following a previously discussed design.[10] The inductors are spiral inductors patterned from the same metal as the ground plane of the antenna array. All the capacitors are parallel plate capacitors that use the same metal-dielectric-metal layers as the antenna microstrip; some are shunt capacitors directly to the ground plane, others are series capacitors defined by patterning the ground plane layer. The filters have large out-of-band impedances so they can be placed in parallel. The transmission of the full network, including the lengths of microstrip from the point where the microstrip branches into four lines to the four filters, has been modeled in Sonnet[§], and the observed performance matches the design quite well; see Figure 5.

### 3.2.2 Microwave Kinetic Inductance Detectors

MUSIC uses microwave kinetic inductance detectors (MKIDs) to detect the optical power exiting each of the bandpass filters. The rationale for using MKIDs in this application is twofold. Foremost, they are highly multiplexable, offering the ability to read out the 2304 detectors required relatively inexpensively and easily. A secondary consideration leading to the use of MKIDs was the interest in demonstrating them in a large-format array with sensitivity close to the background limit.

The MKID device used is shown in Figure 6. The resonator is a hybrid between a coplanar waveguide (CPW) and a lumped element resonator, using the inductive portion of a CPW (6-μm center conductor, 2-μm gap) but expanding the capacitive end into an interdigitated capacitor (IDC). This was done in order to make use of the reduction in two-level system (TLS) noise that had been previously demonstrated.[11] An identical design is used here, with a 10-μm-wide electrode and a 10-μm gap to the ground plane. A larger gap would further reduce TLS noise, but the MKID would become so large that the focal plane fill factor would be reduced substantially. The 10-μm design has an area of 0.7 mm$^2$; with four per antenna, the fill factor is already reduced to 86%, neglecting dead space for feedlines, etc.

---

[§]http://www.sonnetsoftware.com

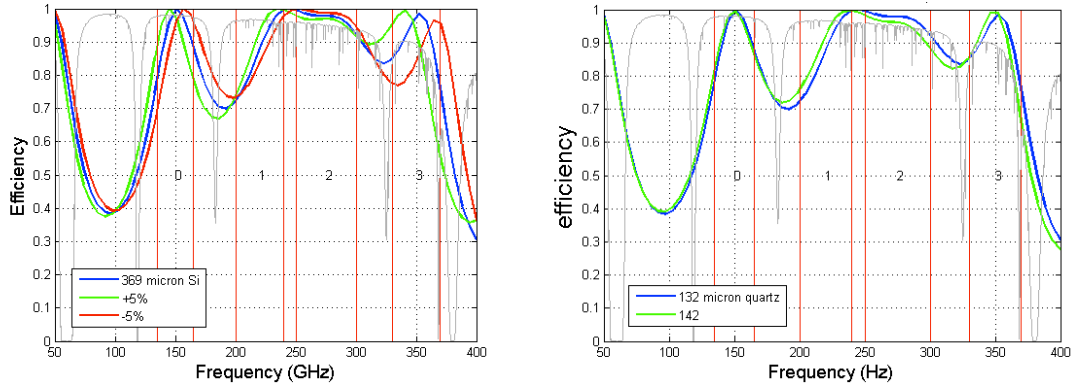

Figure 4: Calculated efficiency of the phased-array antenna as a function of frequency overlaid with atmospheric transmission for PWV = 0.5 mm at the CSO (very good conditions). The calculation is done for a nominal silicon substrate thickness of 369 μm, z-cut crystal quartz antireflection layer thickness of 132 μm, and backshort spacing of 250 μm. The alternate curves in the left plot show the effect of varying the silicon thickness by ±5% and the right plot shows the effect of increasing the antireflection layer thickness to 142 μm. In both cases, the band-averaged effect is quite small, no more than a few %. The calculation is done using the method of moments for an infinite array of infinitely long slots. It accounts for the efficiency of the antenna feed tree, the antenna's radiative efficiency, and the transmission of the Si and quartz layers. It does not account for power lost to substrate modes (these are not present in an infinite arrays); nor to inefficiency due to the finite length of the slots and the corresponding slot impedance variation near the ends. These are both expected to be approximately 10% effects, roughly, and are accounted for in Table 1.

The resonator is fabricated entirely from the antenna ground-plane niobium except for a short, 0.35-mm long, approximately 60-nm thick Al section at the inductive end. This hybrid design was developed in response to observation of direct absorption of power into the MKID in earlier devices;[12] by reducing the Al length, the length of the MKID subject to this direct absorption is proportionally reduced. Power is fed into the MKID from the antenna by routing the 2-μm-wide microstrip exiting a particular bandpass filter over the Al section so that the Al CPW center conductor becomes a lossy ground plane for the microstrip and incoming power is absorbed and converted to quasiparticles. The typical Al film has a sheet resistance of 0.22 Ω/square, yielding an absorption length of 0.6 mm and a calculated absorption efficiency of more than 90% for all the MUSIC mm-wave bands.

The IDC appears to function as an antenna on its own. Power from the IDC section can be transmitted to the Al section along the niobium CPW section, and this was in fact observed in the first IDC resonators illuminated optically by noting that even "dark" resonators (resonators with no antenna feeding them) responded to optical power.[13] A stepped-impedance filter has been added to the Nb portion of the inductive meander to block this radiation. It has been optimized to maximally block radiation in the range 70–450 GHz, covering all frequencies between the nominal pair-breaking frequency of Al and the low-pass edge of the metal-mesh filter (with some margin). It consists of a 24-section filter with consecutive sections varying in width between 14-μm center conductor, 2-μm gap and 1-μm center conductor, 8.5-μm gap and with the lengths of the sections starting at 176 μm and decreasing in a geometric series by a factor 0.94 between each successive section. The parameters of the filter were optimized by calculating the transmission spectrum using Sonnet and then integrating it with a $\nu^2$ spectrum and the metal-mesh low-pass filter transmission spectrum above 70 GHz. The overall length was constrained by the preexisting resonator design (set by the IDC dimensions and the inductor length intended to provide a specific range of resonant frequencies), so the only parameters available were the number of sections (16, 24, and 32 were tried), the length of the first section, and the length reduction factor between consecutive sections. With an earlier version of this filter, it was observed that the remaining direct absorption was approximately proportional to the length of the Al section, indicating this filter was functioning effectively and IDC-coupled radiation was a subdominant contribution. The design was changed to the above in the most recent fabrication runs in order to increase the center-conductor ground-plane gap (from 1 μm to 2 μm): it was feared that fabrication yield problems were due to shorts across this gap arising from incomplete etching.

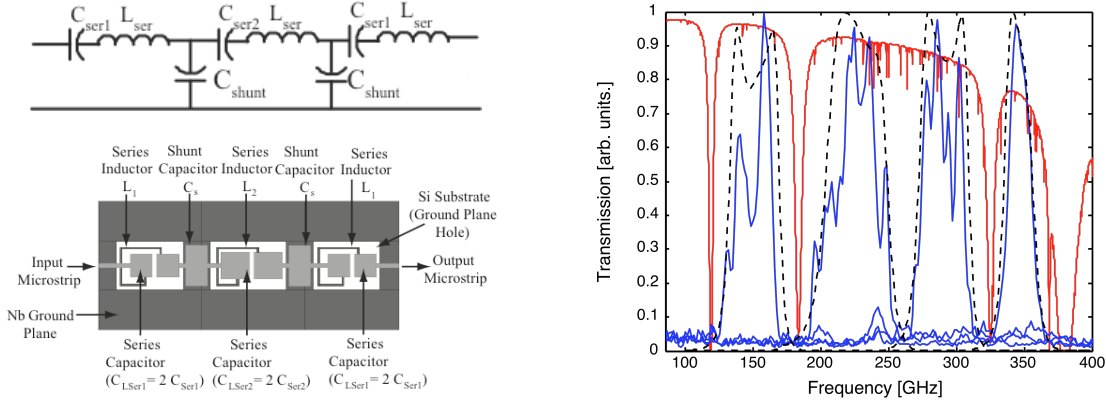

Figure 5: MUSIC bandpass filter design and performance. Left top: Equivalent bandpass filter lumped-element design. Left bottom: Bandpass filter layout. At 272 GHz, the shunt capacitor is 10 μm by 20 μm. The bulk of the metal shown is the ground-plane layer of the antenna. The cutouts allow for the definition of the series spiral inductors and the bottom plates of the series parallel-plate capacitors. Right: Expected (black dashed) and observed (blue solid) transmission overlaid with atmospheric transmission for PWV = 1.5 mm (median conditions) at the CSO (red). The bandpass spectra are individually peak-normalized. They match the atmospheric windows extremely well and show only modest fringing, which is likely due to the dielectric filters.

While the stepped-impedance filter is effective, there remains direct absorption by the aluminum section itself. It has been reduced by shortening the length of the Al section as much as possible from the original 1-mm value to 0.35 mm. Further shortening would result in reflection of mm-wave power from the microstrip. It appears that this coupling is unpolarized, so, to further reduce this direct absorption, a polarizer will be installed just in front of the Lyot stop. It will reflect this undesired polarization to the 4 K shield's inner surface. With the new baffle and this polarizer, the expected power received from a blackbody source via this direct absorption will be 4% to 20% of that received via the antenna, depending on the device and the mm-wave band. The direct absorption remains large because of its broad bandwidth, roughly 10 times the mm-wave bandpass filter bandwidths.

The radiofrequency design of the MKIDs involves the coupling to the feedline and the frequency ordering scheme. In general, the feedline coupling quality factor, $Q_c$, should be designed to match the internal quality factor, $Q_i$, which itself should be set by the quality factor due to quasiparticles under optical load, $Q_{i,qp}$. The expected optical loads and $Q_{i,qp}$ values are given in Table 2, implying an optimal $Q_c \approx$ 100,000–150,000. The original design was for $Q_c \approx$ 50,000 based on a lower early estimate for $Q_{i,qp}$ that was based on too optimistic assumptions about optical efficiency. Sonnet calculations showed this value could be obtained by spacing the MKIDs away from the 20-μm center width, 20-μm gap CPW feedline by 63 μm, where this separation indicates the distance from the outer edge of the CPW feedline gap to the edge of the IDC gap. In practice, however, $Q_c \approx$ 60,000–110,000 is observed, higher than expected. Fortunately, this range turns out to be a good match to the observed $Q_{i,qp}$: the coupling factor $\chi = Q_r^2 Q_c^{-1} Q_i^{-1}$, where $Q_r^{-1} = Q_c^{-1} + Q_i^{-1}$, $Q_i^{-1} = Q_{i,qp}^{-1} + Q_{other}^{-1}$, and $Q_{other}$ refers to other, subdominant forms of loss, is above 0.22 for the observed $Q_{i,qp}$, close to its maximum value of 0.25.

The MKID radiofrequency assignment is given in Figure 6. It is not monotonically increasing with position along the feedline because, in earlier designs, a very clear RF coupling between adjacent resonators was observed, and it resulted in RF resonant modes that were not localized physically on single resonators. The physics of this for a specific MKID geometry is discussed in detail elsewhere.[14] Generally speaking, this coupling can be reduced by increasing the frequency spacing between physically adjacent resonators. Therefore, the final scheme makes use of the fact that each spatial pixel requires four adjacent MKIDs for the four mm-wave bands. The resonant frequencies of the MKIDs in a particular mm-wave band in adjacent spatial pixels are placed adjacent in radiofrequency. Because they are separated physically by the three resonators of the other three mm-wave bands, the coupling between these frequency-adjacent resonators is negligibly low. In this scheme, all the resonators for a given mm-wave band form a single block in RF space, and there are four such blocks for the four mm-wave bands. If $\Delta f_{RF}$ is the radiofrequency spacing between MKIDs that are adjacent in radiofrequency, and there are $N_{pix}^{read}$ spatial pixels in a single readout group, then physically adjacent resonators will be spaced by $N_{pix}^{read} \Delta f_{RF}$. In practice, $N_{pix}^{read} = 36$ was chosen based on previous experience of how closely resonators can be placed in radiofrequency without collisions (resonant frequencies separated by less than a few FWHMs), $\Delta f_{RF}$ = 3–5 MHz, and the available readout bandwidth per readout block, roughly 450 MHz (Section 3.4).

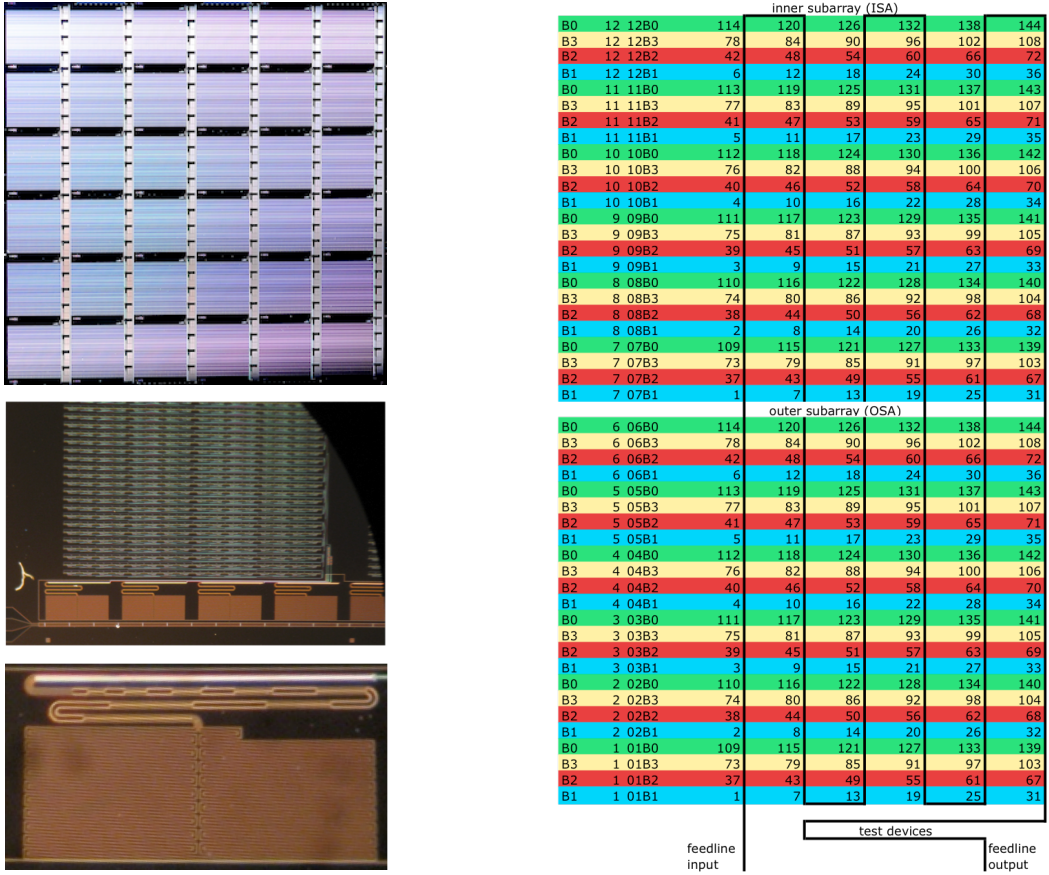

Figure 6: MUSIC focal plane architecture. Left top: Geometry of a 6 × 6 subarray. Each large square is one phased-array antenna spatial pixel. The four small rectangles to the right of each antenna are the four MKIDs for the four mm-wave bands of that spatial pixel. The bandpass filters are too small to be visible. Left middle: Closeup of the four MKIDs for a given pixel (now rotated 90° clockwise). The microstrip emerges from the antenna on the right and enters the bandpass filter block (only three filters are present in the picture). The individual bands' microstrips exit the filters and run to the three MKIDs (on this wafer, the fourth MKID was not optically coupled to the antenna). The readout CPW feedline passes along the bottom of the picture and couples capacitively to the capacitive sections of the resonators. Left bottom: Closeup of a single MKID. The bottom portion is the interdigitated capacitive section and the top portion is the inductive meandered section. The resonator is all niobium except for the bright section at the top, which is Al (1 mm long in this picture, 0.35 mm in the fielded devices). The structure along the inductive meander is the stepped-impedance filter discussed in the text. Right: Pixel physical and frequency layout scheme. The colors indicate the mm-wave band (B0: green; B1: cyan; B2: red; B3: yellow). The numbers indicate the frequency ordering. One can see that resonators that are vertically adjacent in the same mm-wave band are consecutive in frequency. Four vertically adjacent resonators from the four different mm-wave bands form a single spatial pixel. The "inner subarray" and "outer subarray" each consist of a 6 × 6 array of spatial pixels, with the inner array closer to the azimuth midline of the field-of-view, fabricated together as a unit and sharing a single feedline, as described in the text. A single coax input and output serves an entire tile. Test devices to measure the intrinsic α of the Al film and to measure the loss tangent of the dielectric used in the antenna microstrip are placed along the feedline at the bottom right.

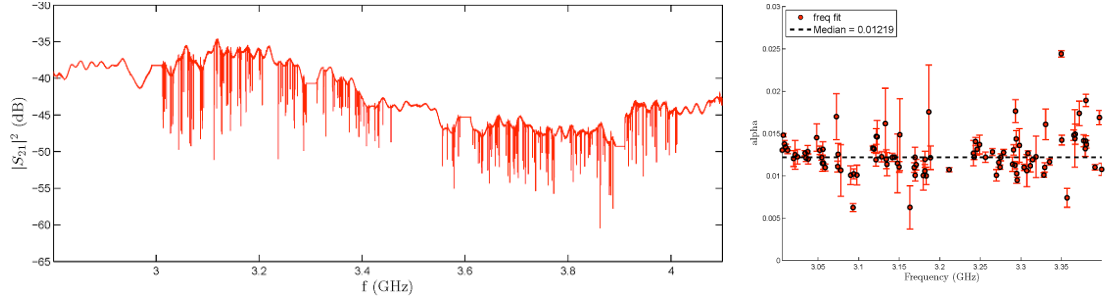

Figure 7: Left: Transmission spectrum of a typical MUSIC subarray, measured dark. The separation between the two 500-MHz wide resonator bands is visible. Right: measurements of α for the lower 500 MHz band of the same device.

The result is $\Delta f_{RF}$ = 3.16 MHz, yielding a spacing $N_{pix}^{read} \Delta f_{RF} \geq 114$ MHz between spatially adjacent resonators.¶ One challenge in such a scheme is that fabrication variations with physical position can result in a position-dependent scaling of the frequency pattern. The roughly 35–40 mm distance between the upper edge of the RF band corresponding to one mm-wave band and the lower edge of the RF band for the next mm-wave band (corresponding to two corners of a 6 × 6 subarray of pixels) may cause the two sets of resonators to intermix to some extent, placing resonators unintentionally adjacent in radiofrequency. Fortunately, such resonators will be physically well-separated and thus will not form coupled modes. Individual pairs of overlapping resonators may be rendered unusable by collisions, but the integrity of the overall frequency scheme will be preserved.

A single 6 × 6 subarray occupies about 500 MHz of RF bandwidth. However, the cryogenic analog RF components, especially the HEMT amplifier, have much larger bandwidths. Thus, to minimize the number of cryogenic readout paths, two such subarrays are fabricated on a single tile and share a feedline with one subarray displaced in radiofrequency from the other by about 600 MHz to avoid RF overlap. This scheme is observed in the $S_{21}(f_{RF})$ plot shown in Figure 7. This scheme also reduces the dead space in the focal plane lost to transitions to/from coax. The two subarrays use separate room-temperature readouts, as described in Section 3.4.

The ground planes on the two sides of the CPW are tied together by 7-μm-wide straps every 400 μm. These inhibit the excitation of slotline modes due to the imperfect grounding of the ground plane.[15] The straps are part of the antenna ground plane layer, while the CPW feedline is fabricated using the microstrip layer. The microstrip insulator serves to insulate the CPW center conductor from the straps.

### 3.2.3 Device Fabrication and Yield

The devices are fabricated on high-resistivity silicon substrates in JPL's MicroDevices Laboratory. Both Nb layers and the Al MKID layer are deposited by DC-magnetron sputtering. The silicon nitride is deposited via inductively coupled plasma-enhanced chemical vapor deposition (ICP-PECVD). All layers are patterned by plasma etching. A stepper mask is used, which is convenient for producing the repeated antenna pattern, allowing for changes in the capacitance of the IDC to vary the resonant frequency. Two tiles fit on a single 100 mm Si wafer.

Yield has proven to be a challenge in full arrays. During the first half of 2012, we fabricated a total of 18 tiles in two batches of 8 and 10 tiles each, with 2 tiles per wafer, counting only wafers that made it all the way through fabrication. In the first batch, yield of tiles with feedline continuity was low, and yield of resonators on tiles with good feedlines was low. In the second batch, substantially more effort was put into cleaning the devices between fabrication steps, and this seems to have improved both feedline continuity and resonator yield. Feedline continuity is likely determined by step coverage: the feedline passes over ground straps every 400 μm, corresponding to roughly 2000 such steps per wafer, and a single failure causes feedline discontinuity, so the step coverage failure probability must be well below 0.1% to ensure continuity. Resonator viability is likely determined by one or both of two failure modes. One is the Nb/Al interface in the MKID, which, if poor, can lead to loss and $Q_i$ so depressed that the resonance is undetectably shallow. This interface is not made in a single vacuum cycle. Rather, the Al is deposited and patterned, requiring exposure to air, then it is ion-milled before the deposition of the ground-plane Nb that also provides the resonator Nb. Another possible failure mode is roughness of the edges of the CPW gap due to imperfect etching, which may also limit the resonator $Q_i$. This gap is etched last, after all other processing, to ensure it is as clean as possible, but defects or residue from earlier processing stages may roughen this edge.

---

¶To avoid the RF transmission dip in the readout near the LO frequency (Section 3.4), 10 MHz on either side of the LO is discarded, causing the separation to increase to 130.5 MHz in some cases.

With the improved yield of tiles from the second batch, the focal plane has recently been upgraded to eight tiles that all show a usable yield of greater than 50% live resonators. Analysis of these new devices is still ongoing and so final yield, resonator performance, and optical efficiency numbers are not available yet.

### 3.2.4 Device Characterization

To determine optical efficiency as well as predict sensitivity, we have developed a detector model based on the BCS theory of superconductivity and the Mattis-Bardeen theory of electrodynamic response of superconductors, as is described in more detail elsewhere,[16] with the following parameters: $\eta_{opt}$, the overall efficiency between power incident on the dewar window and power absorbed in the MKID; $\eta_{ph}$, the efficiency for conversion of incoming photon energy into quasiparticles, $\eta_{ph} \approx 0.58$ for $h\nu \gg 2\Delta$,[17] expected to asymptote to unity as $h\nu \rightarrow 2\Delta$; the resonator film thickness $t$ and area $A$, which give the film volume $V = t\,A$; the quasiparticle recombination constant $R$; the excess power on the device from the dewar, $P_{exc}$, beyond what comes in through the window; the kinetic inductance fraction $\alpha$; the gap parameter $\Delta$; and the intrinsic lifetime $\tau_0$. Calculations indicated that, for typical literature Al diffusion constants appropriate for the observed film sheet resistance, the quasiparticle density is sufficiently uniform over the 0.35 mm Al resonator length to obviate corrections for variations in quasiparticle density. (This was not true for the earlier 1-mm length Al section.)

Dark data are used to measure $\alpha$ and $\Delta$ from the known dependence of the quasiparticle density $n_{qp}$ on bath temperature and the relations between the frequency shift $\delta f$ and quasiparticle quality factor $Q_{i,qp}$ and $n_{qp}$. The $\alpha$ values determined from one subarray are shown in Figure 7. Exposures of the detectors to $T_{BB}$ = 77 K and $T_{BB}$ = 300 K blackbody loads outside the dewar, varying the bath temperature, are used to determine $P_{exc}$, $\tau_0$, and the product $\eta \Delta\nu / t$, where $\eta = \eta_{opt} \eta_{ph}$. The maximum power that can be received from the blackbody loads is assumed to be that for a single-polarization, single-moded antenna, $P_{BB} = k_B T_{BB} \Delta\nu$. The film thickness, $t$, is left as a variable because, while a known 60-nm thickness of Al is deposited, an unknown thickness is etched away unintentionally during processing (Al is relatively easy etched). $\Delta\nu$ is measured separately using Fourier Transform spectroscopy. Literature values for the recombination constant, $R$ = 9.4 µm$^3$/sec, and Al single-spin density of states, $N_0$ = 1.7 × 10$^{10}$ µm$^{-3}$ eV$^{-1}$, are used. The basic equations of interest are

$$Q_{i,qp}^{-1}(T_{opt}, T_{bath}) = \alpha\,\gamma\,\kappa_1(T_{bath})\,n_{qp}(T_{opt}, T_{bath}, \Delta)$$

$$\delta f(T_{opt}, T_{bath}) / f_0 = \alpha\,\gamma\,\kappa_2(T_{bath})\,n_{qp}(T_{opt}, T_{bath}, \Delta)$$

$$T_{opt} = T_{BB} + (P_{exc} / \eta\,k_B\,\Delta\nu); \text{ define } P_{opt} = \eta\,k_B\,T_{opt}\,\Delta\nu$$

$$\eta\,k_B\,T_{opt}\,\Delta\nu / \Delta = R\,t\,A\,(\,[n_{qp}(T_{opt}, T_{bath}, \Delta)]^2 - [n_{th}(T_{bath}, \Delta)]^2\,) + \tau_0^{-1}\,t\,A\,(\,n_{qp}(T_{opt}, T_{bath}, \Delta) - n_{th}(T_{bath}, \Delta)\,)$$

where $\gamma = -1$ under the assumption of the local limit for a thin film, $n_{th}(T_{bath}, \Delta)$ is the thermal quasiparticle density for the given bath temperature $T_{bath}$ and gap parameter $\Delta$, and $\kappa_{1,2}(T_{bath}) = |\,\sigma_{1,2}(T_{bath}) - \sigma(0)\,| / |\,\sigma(0)\,| / n_{qp}$ (where $\sigma_{1,2}(T_{bath})$ are the real and imaginary parts of the complex conductivity) are integrals from Mattis-Bardeen theory. A Markov Chain Monte Carlo is used to perform the fitting to ensure that the full parameter space is explored well and to obtain reliable uncertainty estimates via the posterior distributions. The intrinsic lifetime $\tau_0$ is usually not very well determined, but the analysis is also not very sensitive to it because the recombination time $\tau_{rec} = R\,n_{qp}$ is usually much smaller.

### 3.2.5 Expected Sensitivity

From the above parameters, the expected contributions to the detector's noise-equivalent power (NEP), noise-equivalent flux density (NEFD), and noise-equivalent temperature (NET) can be calculated. One must assume a value for $t$, and it is typically true that NEP will depend on the choice but NET and NEFD will not be very sensitive to it. The expressions are (at signal frequency f, in contrast to radiofrequency $f_{RF}$)

$$\text{NEP}^2(f) = (\,\text{NEP}_\gamma^2 + \text{NEP}_{gr}^2 + \text{NEP}_{TLS}^2 + \text{NEP}_{ampl}^2\,) / (\,1 + [2\pi f \tau_{qp}]^2\,)$$

$$\text{NEP}_\gamma^2 = 2\,P_{opt}\,h\,\nu + 2\,(\,P_{opt}^2 / \Delta\nu\,)$$

| Band | $P_{opt}$ ant [pW] | $P_{opt}$ dir [pW] | $T_{opt}$ ant [K] | $T_{opt}$ dir [K] | $Q_{i,qp}$ [×10^5] | NEP shot [aW/√Hz] | NEP Bose [aW/√Hz] | NEP gr [aW/√Hz] | NEP TLS [aW/√Hz] | NEP ampl [aW/√Hz] | NEP Total [aW/√Hz] | NET Total [mK√s] | NEFD Total [mJy√s] |
|---|---|---|---|---|---|---|---|---|---|---|---|---|---|
| B0 | 0.9–1.5 | 0.6–1.2 | 27–44 | 17–34 | 1.3–1.6 | 18–25 | 13–23 | 20–26 | | | 39–59 | 0.9–1.4 | 39–62 |
| freq | | | | | | | | | 50–80 | 7.4–11 | 59–63 | | |
| diss | | | | | | | | | | 28–47 | 42–63 | | |
| B1 | 2.6–4.1 | 0.6–1.2 | 40–51 | 7–14 | 0.9–1.2 | 31–40 | 22–35 | 28–35 | | | 62–91 | 0.8–1.0 | 37–44 |
| freq | | | | | | | | | 74–110 | 12–19 | 88–130 | | |
| diss | | | | | | | | | | 50–79 | 68–100 | | |
| B2 | 2.4–2.7 | 0.6–1.2 | 60–67 | 14–26 | 1.1–1.3 | 34–39 | 23–30 | 27–30 | | | 63–76 | 1.6–1.8 | 67–78 |
| freq | | | | | | | | | 70–92 | 12–14 | 87–110 | | |
| diss | | | | | | | | | | 47–58 | 68–82 | | |
| B3 | 1.0–2.0 | 0.6–1.2 | 65–93 | 22–46 | 1.2–1.5 | 27–38 | 15–31 | 19–27 | | | 43–69 | 2.5–5.1 | 110–220 |
| freq | | | | | | | | | 47–81 | 7–11 | 60–99 | | |
| diss | | | | | | | | | | 26–48 | 45–74 | | |

Table 2: Median expected optical loads, $Q_{i,qp}$, and sensitivities by mm-wave band for the four well-characterized subarrays (two tiles) displayed in Figure 3. The optical loads received via the antenna and via direct MKID absorption are indicated by "ant" and "dir". The frequency-dependent TLS noise is calculated at a temporal frequency of f = 1 Hz, and TLS noise and amplifier noise are calculated for $P_{read}$ = 10 pW. These estimates use measured dewar optical loading and optical efficiency and make assumptions about loading from the mirrors, telescope, and atmosphere (for median sky conditions, PWV = 1.5 mm). It is assumed that the spillover outside the tertiary mirror can be eliminated by the planned baffle, but the optical efficiency is corrected downward for the elimination of that spillover component of the efficiency. It is also assumed that the polarizer eliminates half of the direct absorption by the Al MKID section. Numbers are given separately for frequency and dissipation direction readout where relevant. The range given is the range of median values over the four well-characterized subarrays. The last two columns give total NETs and NEFDs assuming optimal combination of frequency and dissipation information.

$$NEP_{gr}^2 = t A (\Delta / \eta_{ph})^2 (2 \tau_0^{-1} n_{qp} + 4 R n_{qp}^2 + 4 R n_{th}^2)$$

$$NEP_{TLS}^2 = (t A \Delta / \eta_{ph} \tau_{qp})^2 (\alpha \gamma \kappa_2 / 2)^{-2} S_{\delta f/f0, 1\,Hz}^{TLS} (f / 1\,Hz)^{-1/2} (\chi Q_{i,qp} P_{read} / 1\,pW)^{-1/2}$$

$$NEP_{ampl}^2 = (t A \Delta / \eta_{ph} \tau_{qp})^2 (n_{qp} / \chi r_\kappa)^2 (k_B T_N / P_{read})$$

$$NET = NEP / (k_B \Delta\nu \, \eta_{opt} \, \eta_{tel} \, \eta_{atm})$$

$$NEFD = NEP / (A_{tel} \Delta\nu \, \eta_{opt} \, \eta_{tel} \, \eta_{atm})$$

where the individual NEP contributions are from photon noise (shot and Bose), quasiparticle generation-recombination noise, TLS noise, and amplifier noise, and where $\tau_{qp}^{-1} = 2 \tau_{rec}^{-1} + \tau_0^{-1}$, $P_{read}$ is the readout power on the feedline, $S_{\delta f/f0, 1\,Hz}^{TLS}$ is the TLS noise amplitude at f = 1 Hz and $P_{read}$ = 1 pW, $r_\kappa = 1$ for dissipation direction, $r_\kappa = \kappa_2/\kappa_1$ for frequency direction, $T_N$ is the HEMT amplifier noise temperature, $A_{tel}$ is the telescope area, and $\eta_{tel}$ and $\eta_{atm}$ are the transmission of the telescope and atmosphere. The above formulae give the NEPs for the frequency and dissipation direction with the appropriate choice of $r_\kappa$ and dropping the TLS term for dissipation readout. The expected sensitivities are shown in Table 2. These sensitivities assume that the planned optical baffling eliminates the observed spillover past the tertiary mirror.

### 3.3 Cryomechanical Design

The cryostat and sub-Kelvin cooler were described in detail elsewhere[18] and there have been essentially no changes from that design aside from the completion of the design of and fabrication of the focal plane unit. Please refer to Figure 8. The dewar is cooled by a Cryomech PT-415 pulse-tube cooler, which provides cooling powers of 1.35 W at 4.2 K (PT2) and 40 W at 45 K (PT1). The cooler assembly is mounted to the cryostat via stainless steel bellows. A remote motor is employed and is mounted to the telescope support plate (rather than the dewar itself) to avoid communicating its vibrations to the focal plane. The PT1 and PT2 heads penetrate into the dewar and cool its corresponding stages via flexible thermal straps consisting of copper braid soldered into copper attachment blocks, also intended to reduce vibrations. Where the cryocooler assembly penetrates the PT1 plate, a radiation baffle has been placed to prevent 300 K radiation from reaching the PT2 stage. There is no penetration of the PT2 plate. The PT1 and PT2 plates are made of gold-plated 6061 Al. The attachments between the 300 K vacuum shell, the PT1 plate, and the PT2 plate are made using G-10 tubes of 2-inch diameter and 1/16-inch wall thickness, three for each layer.

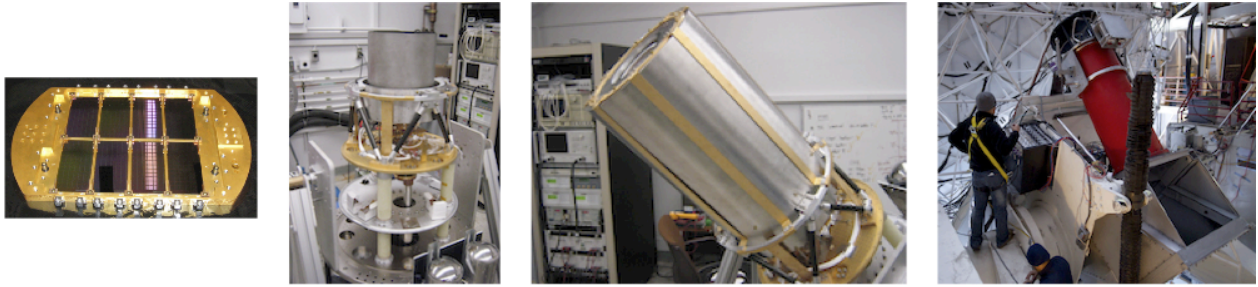

Figure 8: MUSIC cryostat. From left to right: focal plane assembly with 8 tiles installed (niobium backshort removed); the dewar with the niobium magnetic shield installed on the sub-Kelvin IC stage; the optics tube/A4K magnetic shield containing the focal plane components installed on the PT2 stage of cryostat; and the cryostat and its relay optics mounted to the telescope along with one of two readout electronics crates.

The dewar's radiation shields are made from 1100-O Al tube welded into 6061 Al flanges; the former provides excellent thermal conductivity, while the latter is harder. The top plates of the shields, where the various thermal blocking filters sit, are 6061 Al plate. A multilayer insulation blanket consisting of 20 layers of 0.001-inch aluminized mylar spaced by layers of spun polyester is attached to the outside of the PT1 shield. The aforementioned metal-mesh IR shaders and PTFE thermal blockers are attached to the top plates of the radiation shields. The PTFE filters are heat-sunk by copper heat straps.

A Chase Cryogenics $^3$He/$^3$He/$^4$He closed-cycle sorption cooler (CRC10-023) is mounted to the PT2 plate. Its "ultracooler" (UC) stage provides a cooling power of 3 μW at 240 mK and its "intercooler" (IC) stage provides 100 μW at 350 mK. A full cycle of the refrigerator requires roughly 6 hours from start to stable base temperature. The hold time is well in excess of 24 hours and may permit two nights operation per cycle when fully optimized, limited by the $^3$He charge of the IC. It is necessary to turn off the HEMT amplifiers (Section 3.4) during the refrigerator cycle to ensure sufficient $^4$He is condensed.

The mechanical structure inside the PT2 radiation shield consists of an optics tube/magnetic shield surrounding the focal plane hardware. The details of the assembly are described elsewhere.[18] The entire structure is supported from a truss composed of an Al ring attached to a set of carbon-fiber reinforced polymer (CFRP) rods to ensure mechanical stiffness in response to telescope motions; they are designed so that the G-10 suspension rods at PT1 and PT2 dominate the system's flexure. The Al ring is roughly 200 mm above the PT2 plate. It is thermally sunk to the PT2 plate via copper straps. A 2-layer cylindrical Amuneal A4K magnetic shield mounts to this 4 K ring. The A4K shield has two circular penetrations at the bottom for wiring and heat straps, one at the top for the optical path, and four on the sides for mechanical attachments to enter for the optics tube. The optics tube is an Al tube to which the cold lens, Lyot stop, and 420 GHz metal-mesh low-pass filter mount. A cylindrical optical baffle extends from the lens mount toward the Lyot stop. All the inside surfaces of the optics tube and the baffles are blackened with Stycast loaded with carbon[6] and stainless steel.[7,7] As noted earlier, this structure appears to be insufficiently absorbing, so an additional set of optical baffles between the lens and the Lyot stop will be installed. A polarizer will be installed on the sky side of the Lyot stop to reflect the unused polarization to the inside of the 4 K radiation shield and thereby reduce direct absorption.

The focal plane hardware is suspended from the optics tube. CFRP legs are again used to form trusses to suspend the IC stage from the optics tube and the UC stage from the IC stage. A niobium magnetic shield resides at the IC stage, consisting of a half-open cylinder extending from below the focal plane up to the lens. It has two penetrations at the bottom for wiring and heat straps. The UC stage consists of a simple plate to which the focal plane unit is attached. Stiff copper heat straps extend from the UC and IC stages through the shield penetrations, and the final connections to the sub-Kelvin cooler are made with flexible thermal straps consisting of 15 (IC) or 20 (UC) layers of 0.004-inch thick ETP copper foil that are e-beam welded into end blocks, which are gold-plated and attached with many large screws. In addition, a small satellite IC-cooled stage, stood off from the PT2 plate via a CFRP truss, serves to heat sink the readout coaxes and wiring prior to entering the magnetic shield/optics tube and going to the focal plane.

The focal plane unit consists of an OFHC copper plate with cutouts to accept the detector tiles and their anti-reflection tiles. The stackup is held in place by beryllium-copper spring clips. The 576 spatial pixels are divided into 8 individual tiles of 72 spatial pixels each, with one feedline per tile. Duroid circuit boards around the outer edge of the detector region provide electrical contact to the feedline and RF grounding of the detector tile ground plane. They carry 16 blind-mate coaxial connectors that mate to the RF coaxes that are mounted to the UC stage. This design allows for the focal plane unit to be dropped in blindly from above without any need to tighten coaxial connectors. This is necessary given the difficulty of accessing the focal plane unit inside its surrounding shields.

Gold wirebonds are made from a border of gold metal film on the detector tiles to the focal unit copper in order to ensure adequate heat sinking of the substrate in the face of residual thermal infrared radiation. The film is 350 nm thick, 1 mm wide, and is complete on three sides of each tile. The wirebond density is roughly one per 0.5 mm. Calculations indicate that this wirebond density is required for the conductance from the substrate to the focal plane copper to not be limited by the bonds, but rather by either Kapitza conductance between the silicon substrate and the gold film or by transverse conductance through the gold film to the wirebonds. Measurements of dielectric loss tangent test devices on the tiles (Figure 6) indicate the silicon wafer heats up by no more than a few mK above the bath temperature with this heat sinking in place.

### 3.4 Readout Electronics and Data Acquisition

The MUSIC readout electronics make use of the key design convenience of MKIDs, which is that they largely eliminate cryogenic readout components in favor of room-temperature electronics. The architecture of the readout has been described in great detail elsewhere,[19] and the final version will be described in a future paper,[20] so only a summary and substantive changes are presented here.

The component of the readout electronics in the dewar consists of the coaxial cable runs and cryogenic low-noise amplifiers. Stainless steel coaxes, 0.085 inches in diameter, are used for the run from 300 K to PT2, with clamp blocks at the PT1 stage. These coaxes terminate in 20 dB attenuators at the PT2 plate. The path from the PT2 plate to the focal plane and back to the PT2 plate uses 0.064-inch diameter NbTi coaxes heat sunk at the IC stage using clamp blocks and 10 dB attenuators on the input side and 1 dB attenuators on the output side. The exiting NbTi coaxes then mate to 0.5--11 GHz HEMT amplifiers on the PT2 plate, supplied by S. Weinreb. These HEMTs have noise temperatures of 3 K to 4 K in the 3–5 GHz range of interest for MUSIC. The HEMTs were tested individually using a separate 4 K test dewar with a cryogenic thermal load. The HEMT bias points were established in this setup by first minimizing the noise temperature, which tends to yield a broad plateau in bias parameters, and then by maximizing the gain within that region. Typical drain voltages and currents are 1.5 V and 30 mA ($\approx$ 45 mW power dissipation), while the optimal gate voltages vary from 0.5 V to 2.1 V. Gain slopes as a function of gain and drain voltages, $\partial G/\partial V_g$ and $\partial G/\partial V_d$, were found to have typical values of 300–600/V, though values as low as 100/V and as high as 1000/V were found.

The core of the readout electronics outside the dewar is a set of three boards that provide the frequency comb to drive the MKIDs and that receive and demodulate it to recover the astronomical signal: a ROACH signal processing/interface board, an ADC/DAC board, and an IF board. A block diagram and a picture of one readout chain is shown in Figure 9.

The first element is the CASPER[£] Reconfigurable Open Architecture Computer Hardware (ROACH) board, which provides a Virtex-5 FPGA, PowerPC CPU, on-board memory (1 DDR DRAM and 2 QDR SRAM), two fast, wide-format Zdok input connectors, and ethernet interfaces at 10 Gbps, 1 Gbps, and 100 Mbps.

The second component is a custom ADC/DAC board built around the TI ADS5463 500-MHz 12-bit ADC and the TI DAC5618 1000-MHz 16-bit DAC, developed as part of a larger collaboration.[21] These components, in particular the TI ADS5463, were chosen because they provide sufficient SNR (>59 dBFS) to ensure the readout would be HEMT-noise-limited under normal operating conditions ($P_{read} \sim P_{opt} \sim$ 3–30 pW). Each board has two DACs and two ADCs, providing in-phase ("I") and quadrature ("Q") components so that the full sampling bandwidth of the DACs and ADCs can be used for carriers. This board mates to the ROACH board via the Zdok connectors. An earlier version of this board was reported on before.[19] In a more recent revision, greater care was taken with the goal of reducing low-frequency noise arising from temperature-dependence component gains. In particular, the ADC and DAC now reside on the same

---

[£]http://casper.berkeley.edu

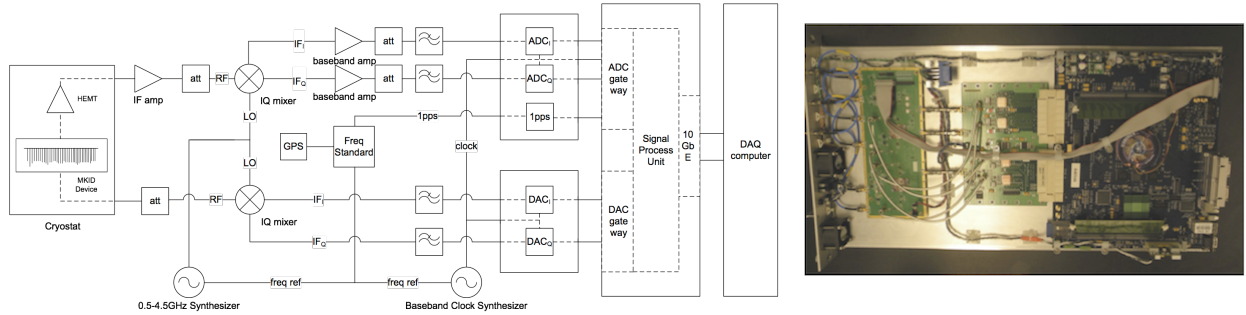

Figure 9: Left: Block diagram of readout system. Right: A single readout assembly consisting of (left to right) an IF board, an ADC/DAC board, and a ROACH board.

physical circuit board and use the same voltage reference, and additional heat sinking has been incorporated to ensure that the limiting thermal impedances are those within the ADC and DAC integrated-circuit packages.

The final component is a custom intermediate-frequency (IF) board that converts from the ADC/DAC baseband (approximately 0–500 MHz; the DAC is run at 491.52 MHz) to the RF band of the resonators (3–4.2 GHz). This board is a new addition relative to the packaged-component version described earlier.[19] It has an on-board frequency synthesizer that supplies the local oscillator (LO) for the upconverting and downconverting IQ mixers. The synthesizer is locked to a 10 MHz rubidium frequency standard, which itself is locked to GPS via a 1 PPS signal. On the generation side, the signal exiting the DAC encounters a transformer and an anti-alias filter followed by the IQ mixer and finally a programmable attenuator to set the power level entering the dewar. On the reception side, additional post-dewar amplification is provided, followed by variable attenuation to ensure the power level entering the mixer is kept constant as the input attenuation is varied. After the IQ mixer comes a baseband amplifier followed by further attenuation and finally an anti-alias filter prior to transformer coupling to the ADC. The anti-alias filter and the transformer eliminate 10 MHz of RF bandwidth on each side of the LO and 10 MHz at the each edge of the RF band, reducing the usable bandwidth to about 450 MHz.

The FPGA firmware has three components. The first serves a lookup table from memory (BRAM on board the FPGA for fast access) to the DAC for playback. The buffer is $2^{16}$ samples long in order to be synchronized with the FFT of the received data. The second is a single-stage, $2^{16}$-sample FFT. Because the FPGA clock runs at half the speed of the ADC/DAC clock, the FFT is pipelined, bringing in 2 samples from the ADC per FPGA clock sample and processing them in parallel using standard algorithms. The third component processes the output of the FFT. It selects the FFT bins in which carrier tones are present (based on previous knowledge of the DAC buffer) and low-pass filters and decimates them from the FFT dump rate (precisely 7.5 kHz, corresponding to an ADC/DAC sample rate of 491.52 MHz) to the final desired data rate (100 Hz). It creates data packets that the ethernet server on the PowerPC makes available to a DAQ computer. The output data streams are in "electronics IQ": they are the real and imaginary parts of the output of the FFT, low-pass filtered and decimated. Note that these are in general linear combinations of the I and Q signals generated at the DAC and of frequency and dissipation signals at the resonator. The firmware can demodulate and decimate 192 carriers.

Great care was taken in the mechanical, electrical, and thermal infrastructure for these readout boards, understanding that 1/f stability was paramount. In addition to liberal use of ground planes and heat-sinking on the boards, the three boards are mounted to a single, 0.25-inch-thick Al plate, with screws connecting their ground planes to the plate. This large thermal mass provides a high thermal heat capacity to damp thermal fluctuations as well as a large area for air cooling. It also ensures that the interboard electrical connections remain stable and do not modulate the system gain. Eight of these assemblies slide on rails into a crate that has the outside dimensions of a 19-inch rack. Each board has its own front panel, rigidly attached to the Al plate, that carries the input and output RF connectors as well as connectors for the 10-MHz and 1-PPS reference signals, optional external clock input and output, LEDs to indicate power and status, and a power/reboot switch. Each assembly's ROACH board interface connectors (USB, 9-pin serial, and ethernet) are presented at the back panel of the crate through cutouts. The PowerPCs run a linux operating system installed on USB drives attached to the USB connectors. The crate also serves to distribute power from a remote power-supply crate, with independent supplies for the ROACH boards, ADC/DAC boards, and IF boards to ensure the critical analog components are not disturbed by power supply noise created by the digital systems. Both crates incorporate ample fans to bring air in through the front panel and exhaust it via the rear panel. A single crate with 8 modules dissipates roughly 550 W and its power supply crate dissipates 300 W.

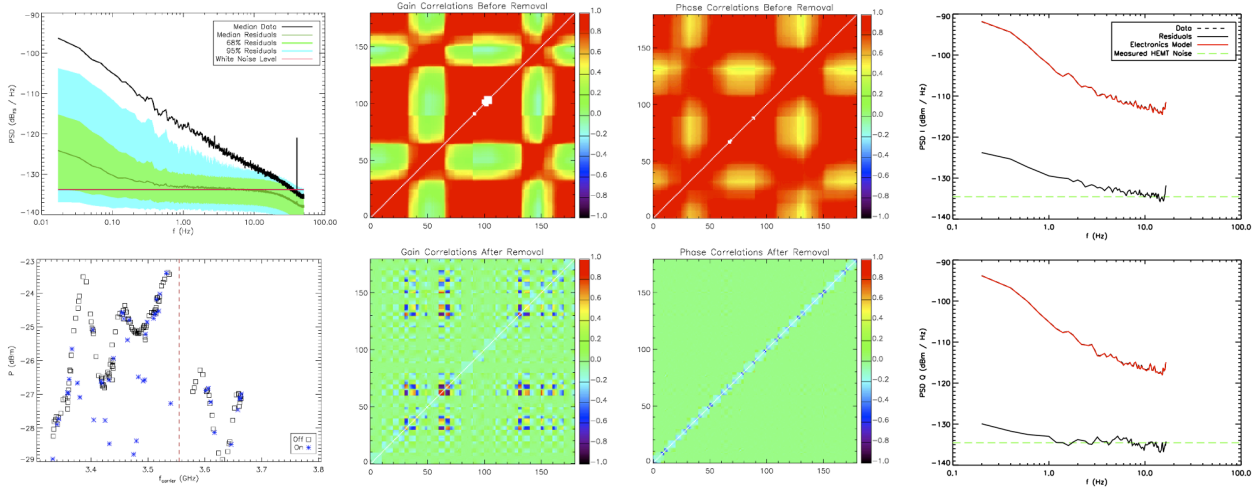

Figure 10: Top left: Noise power spectral density $(I^2 + Q^2)^{1/2}$ for off-resonance tones in laboratory data with the dewar stationary. The thick black line indicates the raw data, the thin green line indicates the median post-removal data, and the green and cyan bands indicate 68% and 95% ranges for the post-removal data. The horizontal red line indicates the inferred white noise level, which should be dominated by HEMT white noise. The expectation for the true HEMT noise level is somewhat uncertain due to uncertainty in the system gain from the HEMT to the ADC, but is in the range of −140 to −135 dB$_{FS}$/Hz, consistent with measurements. Bottom left: Carrier power received at the ADC as a function of radiofrequency for the carrier set used to generate the top left plot. The large variations are not in the programmed carrier power, but are due to the varying RF transmission of the system. Off- and on-resonance tones are indicated; on-resonance tones obviously suffer additional attenuation by the resonators. Center: pre- and post-removal gain and phase correlations among off-resonance tones. The patchy nature of the pre-removal correlations seems to be due to steep variations in the RF gain. To clean a particular carrier, an electronics noise template is built from the 10 nearest carriers that are also in the same "correlation block" observed in the pre-removal correlations. The post-removal correlation plot indicates this works quite well. Right: Noise power spectral densities (I and Q separately) for off-resonance carriers, pre- and post-removal, from on-sky data on the telescope, using the same technique as employed for the top left plot. The pre-removal data and the electronics noise template lie on top of each other. The dashed green line indicates the white noise level inferred from the fit to the correlated noise model; it is quite consistent with the number inferred from lab measurements and with expectations for the HEMT noise. The removal works comparably well as for lab data near 10 Hz, but is less effective below 1 Hz. This degradation remains to be understood, but may be due to motion of the dewar-to-crate coaxes during telescope motion.

The electronics crates are mounted to the CSO's elevation plate, a very short distance away from the dewar itself, to ensure minimal loss in the coax run between the crate and the dewar. Sixteen readout assemblies reside in two crates. Pairs of readout assemblies are joined together using two-way power combiners and dividers so that only 8 readout chains are required in the dewar. Given the large power dissipation, the crate exhaust is ducted away from the telescope primary mirror. The power supply crate sits well way from the electronics crate due to space availability on the telescope and to minimize the heat dump under the primary mirror. The power supply voltages are corrected to account for voltage drop in the long cable run.

In spite of the extensive efforts to reduce 1/f noise, a substantial level remains present. Fortunately, this noise is well-correlated among carriers that are nearby in frequency. By adding a large number of off-resonance carriers interspersed through the readout band (available due to the difference between the number of good resonators per readout block and the 192 tones each readout block is capable of handling, so at least 48 off-resonance tones and in general more), it can be removed effectively. The performance of the electronics is illustrated in Figure 10.

The 16 readout assemblies are connected by 1 Gbps ethernet via a network switch to a central data acquisition computer running linux. A MATLAB[ϵ] program undertakes the main data-acquisition loop. It spawns Python[fi] clients that

---

[ϵ] http://www.mathworks.com

initialize the IF boards to generate the LO and FPGA clocks, load the DAC buffers into FPGA memory, and specify the FFT bins for which data is exported. The MATLAB program then spawns Python clients to read the data being served over ethernet by the individual readout assemblies. The MATLAB loop monitors software-logic signals from the telescope to determine when the observer would like to initiate a new observation, and it has a software-logic interlock with the telescope so that device characterization data can be taken when desired prior to starting an observation (see below). It also acquires pointing and slow-control data from the telescope over ethernet. When an observation (10–20 minutes of contiguous data on a single source) is complete, the MATLAB program initiates merging of the data from the individual readout blocks together with pointing and slow control data to form a single file that contains all relevant data for a given observation, including device characterization data. The file output is netCDF[¿], which is a flexible, self-describing format, chosen also for compatibility with the Bolocam data analysis pipeline[¡].

Prior to each observation, a device characterization is done as noted above because the MKIDs' resonant frequencies move in response to changes in optical loading, which can occur as a source's airmass changes or one slews to a different position in the sky. First, an archival DAC buffer is loaded. The IF board LOs are commanded to step through a range of a few hundred kHz, taking data at each LO setting, to obtain $S_{21}(f_{RF})$ through the resonances. The $S_{21}$ data are analyzed and new resonance frequencies are chosen by finding the maximum in $|dS_{21}/df_{RF}|$ for each MKID. A new DAC buffer is generated with these new resonance frequencies (but with readout power levels unchanged) and loaded into the FPGA. A second LO sweep is done with the updated buffer to obtain the RF transmission for the settings that will be used for data acquisition. The data are stored and then normal data-taking is initiated using the updated buffer.

### 3.5 Data Analysis and Noise Removal

The single-observation netCDF files output by the data acquisition are copied to a different linux machine and processed using an upgraded version of the publicly available Bolocam software pipeline. In addition to the standard functions of such a pipeline—relative gain correction, removal of common-mode sky noise, mapmaking—the pipeline must undertake some MKID-specific functions.

First, the pipeline must determine the various critical directions in the "electronics IQ" plane of the received data: the frequency and dissipation directions as well as the "quasiparticle" direction, which is the linear combination of frequency and dissipation changes that occur when quasiparticles are created by optical power or thermal effects, with slope $\kappa_1/\kappa_2$ relative to the frequency direction. In principle, the LO sweep data and sky noise can be used to determine the frequency and quasiparticle directions, respectively. However, in practice, the LO sweep data may be insufficient due to relatively coarse sampling in frequency (too fine a sampling would require too much time to acquire the data, and carrier frequencies can only be set to 7.5 kHz precision due to the FFT length), and conditions may change by a large enough amount during a single observation that the direction calibration is not valid for the entire observation. In addition, sky noise is not always sufficiently dominant over other noises to provide a precise enough estimate of the quasiparticle direction. "On-the-fly" calibrations will likely be implemented to improve the situation. A chopped mm-wave blackbody source can be directed into the optical train to provide a small (sub-fW power levels), well-defined signal indicating the quasiparticle direction. The firmware can be modifed to frequency-modulate the carriers,[22] which would provide a continuous measurement of the frequency direction. Care must be taken with the latter to ensure that the readout power is not modulated and the quality factor does not change appreciably during the modulation. Additional amplitude modulations of the DAC I, DAC Q, and DAC I and Q may also be included to provide additional monitoring of electronics gain and phase fluctuations.

Second, the pipeline must simultaneously remove electronics noise and sky noise. The challenge of doing so is illustrated in Figure 11. An iterative technique is being employed: while the electronics noise template is independent of the active resonator data because it is calculated using only off-resonance tones, the sky noise template must be determined from the resonator data, and so the sky noise template should be refined as the electronics noise and sky noise correlation coefficients are estimated. Nevertheless, removing these noises has proven challenging because their spectra and amplitudes appear to be sufficiently similar that the assumption of statistical independence of the two signals fails. More sophisticated techniques, such as calculating correlation coefficients using only ranges of audio frequency where one or the other dominates (e.g., sky noise is dominant below ~ 1 Hz, while electronics noise dominates at higher

---

fi http://www.python.org
¿ http://www.unidata.ucar.edu/netcdf/
¡ http://www.cso.caltech.edu/bolocam/AnalysisSoftware.html

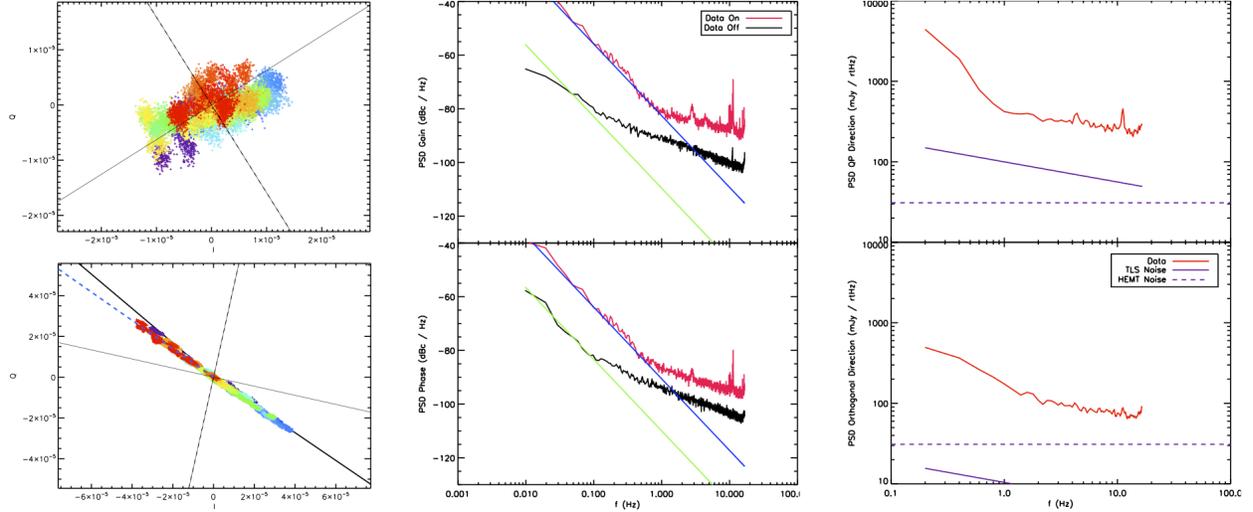

Figure 11: Observed noise behavior. Left: noise for off-resonance (top) and on-resonance (bottom) carriers in "electronics IQ" plane. The mean value over the observation has been removed, moving the data to the origin. The scale is in fractional change relative to the mean $(I^2 + Q^2)^{1/2}$ value. Both panels are isometric but different in overall scale. The scattered points show the data, with the colors indicating independent 5-second sets of data (with gaps in between). In both plots, the thin lines indicate the electronics gain and phase directions (radial and azimuthal directions in the IQ plane). In the lower plot, the thick solid line indicates the data from the sweep of the LO frequency prior to taking the observation; it traces out the resonance curve and should indicate the frequency direction. In this particular plot, decreasing frequency (increasing optical power) is to the upper left. The blue dashed line indicates the "quasiparticle" direction, the linear combination of frequency and dissipation changes that corresponds to optical power deposition. The low-frequency electronics noise is visible in the movement of the noise distribution from scan to scan in the upper plot. The additional noise present in on-resonance tones is visible in the lower plot; the bulk of this noise is along the quasiparticle direction, indicating it is varying optical power from atmospheric fluctuations (sky noise). It is clear that sky noise is dominant but that electronics noise causes deviations from the sky noise direction; separating these two effects has proven challenging. Center: Raw noise power spectral density observed in a particular resonator along with the electronics noise template determined from nearby off-resonance tones. The similarity of the high-frequency slopes of the two spectra strongly suggests electronics noise dominates above about 1 Hz. However, one would expect the off- and on-resonance noises to be equal in these units, dBc/Hz; this discrepancy remains unexplained. The solid lines indicate the expected slope for sky noise, with exponent −8/3. This may be indistinguishably similar to "drift" (PSD $\propto 1/f^2$), as suggested by the match of the low-frequency electronics noise observed in the phase direction for the off-resonance tones to this expected slope. Right: raw on-resonance noise calibrated to flux density units. The noise is below 400 mJy Hz$^{-1/2}$ above 1 Hz. Given the substantial improvement in electronics noise that can be obtained for off-resonance tones, a large improvement in the resonator noise is expected once sky and electronics noise removal are done well. Note also that TLS and HEMT noise contributions will be reduced as new baffling reduces excess optical loading, which degrades resonator responsivity.

frequencies) and accounting for the covariance between the templates when determining removal coefficients, are being developed. This is work in progress that will be reported on in the future.

## 4. CONCLUSION

We have developed a new mm/submm-wave camera, MUSIC, that employs an entirely photolithographic focal plane with a range of new technologies—slot-dipole phased-array antennas, lumped-element on-chip bandpass filters, and microwave kinetic inductance detectors—to obtain an unprecedented combination of spatial and spectral coverage in a single instrument. The instrument is in the midst of being commissioned, with 25% of the focal plane well-characterized and the remainder being characterized. The optical performance largely meets expectations with the exception of distortion, stray light, and optical efficiency. The first can be measured and corrected for, the second will likely be reduced by improved baffling, and the third may be amenable to future fabrication improvements. Device yield has been a challenge but has improved in later device runs and will likely continue to improve in the future. The cryogenics, electronics, and data acquisition systems are performing well. The analysis of the data has proven challenging, but a combination of additional calibration techniques as well as more sophisticated analysis techniques will likely yield much improvement. MUSIC will be available for science observations near the end of 2012. A complete exposition of instrument design and performance will be forthcoming.

## ACKNOWLEDGMENTS

MUSIC was designed and constructed with support from NSF/AST-0705157, NSF/AST-1009939, the Gordon and Betty Moore Foundation, and the Caltech Submillimeter Observatory, which has been supported by NSF/AST-0540882 and NSF/AST-0838261. JS and MIH were partially supported by NASA Postdoctoral Program fellowships. JAS and NGC were partially supported by NASA Graduate Student Research Program fellowships. We acknowledge Kathy Deniston and Barbara Wertz for administrative support and the staff and day crew of the CSO for assistance during design, installation, and commissioning.